# Models of the Mass-Ejection Histories of pre Planetary Nebulae, IV. Magnetized Winds and the Origins of Jets, Bullets, and FLIERs

Bruce Balick[1], Adam Frank[2] and Baowei Liu[3]

[1] Department of Astronomy, University of Washington, Seattle, WA 98195-1580, USA
[2] Departments of Physics and Astronomy, University of Rochester, Rochester, NY 14627, USA
[3] Center for Integrated Research Computing, University of Rochester, Rochester, NY 14627, USA
E-mail: balick@uw.edu

**Abstract**

The influences and consequences of toroidal magnetic fields in shaping the visible lobes of pre planetary nebulae ("prePNe") are explored in this, the last of a series of papers of parameter studies of prePN evolution. To probe these influences we start with the steady, diverging, and field-free wind model of our previous papers and add weak to moderate toroidal fields to the winds in order to generate arrays of outcomes after 500 y, after which the structures grow almost homologously. As expected, toroidal fields in the stellar winds invariably form very thin and dense axial features whose structure is best described as a thin cold jet with an ultra-dense and neutral leading knot, or "bullet", at its tip. The speed of the leading knot depends only on the density contrast (the ratio of injected to ambient gas densities at the nozzle) and wind injection speed, but not on the field strength or opening angle. The lobes formed by the ram pressure of the winds take a variety of forms and sizes that depend primarily on the geometric structure of the injected gas and the density contrast. About 20% of the HST images of prePNe show the unique signatures of shaping by toroidal fields. Pairs of low-ionization knots seen along the major axis of fully ionized PNe, often called "FLIERs", are easily explained as the very dense, cold, and neutral remnants of magnetically formed knots.

## 1. Introduction

This is the fourth publication in a series of studies that trace the evolution of bilobed pre planetary nebulae ("prePN") whose images were captured by the Hubble Space telescope ("HST"). The primary purpose of this paper is to investigate the key signatures and early influences of toroidally magnetized winds emanating from the atmosphere of a central star in the shaping PrePNe represent the adolescence of a planetary nebula ("PN") as it takes its ultimate form. The shaping starts as the mode of mass loss from an AGB star changes from an isotropic "slow AGB wind" for thousands of years (Höfner and Olofsson 2018A&Ar v..26....1H) into a collimated form. This transition can be abrupt, for example as a companion tidally strips the atmosphere of an AGB star. Over the next $\approx 10^3$ y, a collimated flow from the nucleus is supersonically injected into the remaining slow AGB winds at speeds in excess of 100 km s$^{-1}$. After a few hundred years an observable prePN, consisting of displaced and reshaped AGB winds, is fashioned by the ram pressure of the collimated flow and, possibly in some cases, by a magnetic field embedded in the stellar outflow at its source.

Over the past 20y there have been many suggestions that the symmetric shapes of observed bilobed—or "bipolar"—prePNe are shaped by a combination of fast collimated stellar winds with (perhaps) toroidal magnetic fields (e.g. Balick & Frank, 2002ARA&A..40..439B). The evidence for shaping prePNe by such winds is now overwhelming (e.g. Sahai et al., 2011AJ....141..134S); however, the influence of fields still remains unsettled. We focus on the latter in this paper.

In Balick et al. (2017ApJ...843..108B, hereafter "Paper I") and Balick et al. (2018ApJ...853..168B; hereafter "Paper II") we presented numerical non-magnetic hydrodynamic ("HD") models that nicely account for the structure and dynamics of two specific PNe of rather unusual and somewhat complex shapes, OH231.8+04.2 and M2–9, respectively. Our most recent paper, Balick et al., 2019 (hereafter "Paper III"), presented the set of state variables for a rather generic, or "baseline", model for the most common bipolar prePNe: those with pairs of flame-shaped lobes but no signs of a magnetic field. We considered how the choice of state variables and the flow geometry produces realistic outcomes, and we made some general remarks on the geometry and dynamics of the fast collimated outflow that are essential for interpreting observations. (Some highlights will be mentioned later.) We also touched very briefly on the influences of toroidal magnetic fields embedded in the outflows that might be the result of accretion disks at the base of the flows. We concluded that fields are of little relevance for the vast majority of prePNe, but that they may well play an active role in those prePNe that show signs of protrusions or jets along the flow symmetry axis and beyond the lobes. Several examples of candidates for magnetic shaping are shown in Fig. 1.

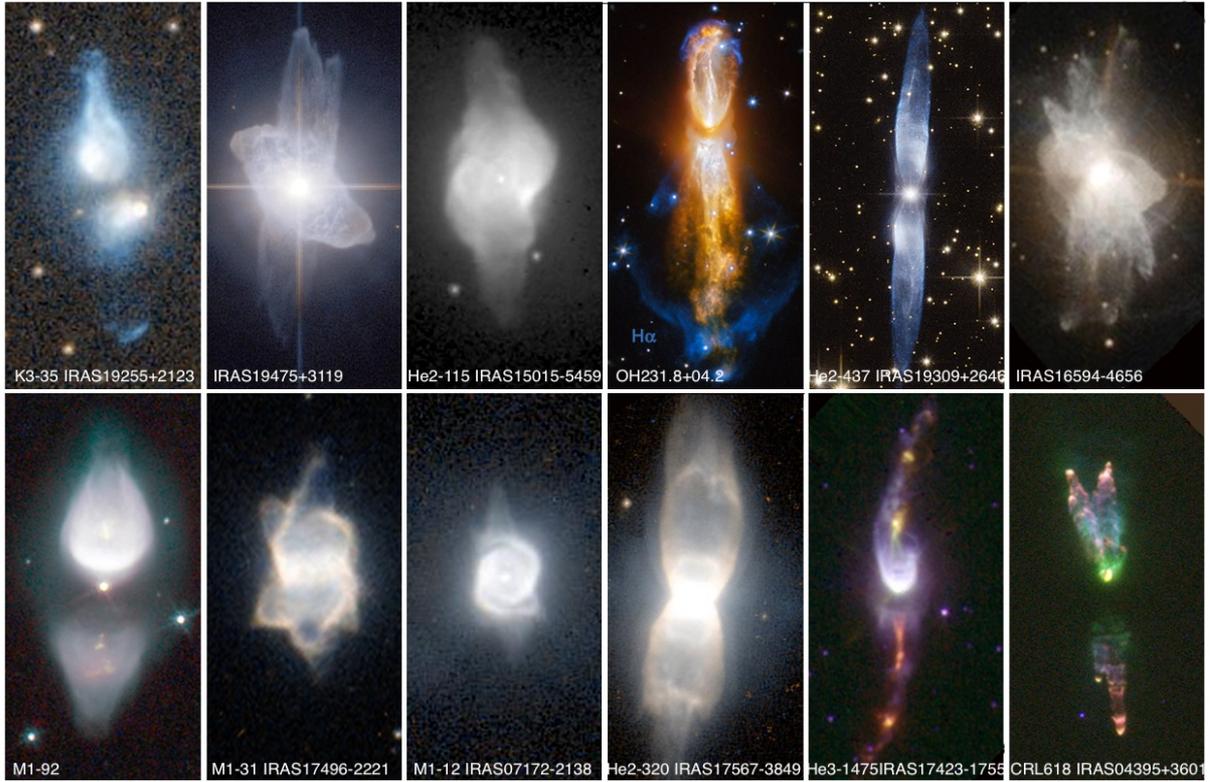

Fig 1. Hubble Space Telescope ("HST") images of a subclass of very compact prePNe that are good candidates for magnetic shaping;; i.e., they contain conspicuous protrusions and shocked knots along their symmetry axis. All images were obtained from sundry public internet sites. Spatial scales, colors, and orientation vary among the images.

All shaping models with toroidal magnetic fields have one very important outcome in common: jets along the field symmetry axis that are collimated by hoop stresses. Therefore the appearance or inference of a jet from an image of an outflow does not, by itself, reveal any additional information about its origin, its history, or its evolution. In this paper we include the influence of toroidal fields with other parameters such as the density and geometry of the flow in order to produce arrays of simulated shapes. Matching observed and simulated shapes is one—and perhaps the only—way to probe the historical evolution of magnetically shaped outflows.

All previous models are anchored to specific, and therefore anecdotal, inflow paradigms. Most share the same mass injection paradigm in which a rotating AGB or white dwarf creates surface winds proposed by Bjorkman & Cassinelli (1993) and developed further by Ignace et al. (1996ApJ...459..671I and 1998ApJ...505..910I). At the time those papers were published the importance of a different and presently favored scenario of mass injection from close binaries with an AGB star—a complex process still being explored (e.g., Zho et al., 2019 (preprint); Kim et al., 2019ApJS..243...35K; Chen et al., 2018MNRAS.473..747C, MacLeod et al., 2018ApJ...868..136M)—had not yet garnered much attention. Generalizing from these highly anecdotal simulations (computed at poorer spatial resolution than ours) is difficult.

We make no assumptions about the structure of the winds from the central core. As a result, the present simulations are more generic and wide ranging than their predecessors.

We start by summarizing the results of non-magnetized inflows (Paper III) whose outcomes nicely match HST images of many prePNe and, when measured, their kinematics. A fairly generic "baseline model" of a non-magnetized flow emerged in that paper[1]. Here we add a toroidal magnetic field $B_z$ injected into base of the flow to the other parameters of the baseline model. We trace the trends of lobes shapes and sizes as the field intensity is increased from weakly to strongly influential.

In section 2 we present of our methodology, describe the general nature of the injected flow and its environment, and show how an array of outcomes vary as flow state variables are changed. We also explore the early years of the flow when the simulated PNe were imprinted with their later shapes. In section 3 we will compare our results to images of preNe and PNe, including the poorly understood low ionization knots FLIERs. We also contrast our outcomes with previous MHD simulations of prePNe. We end with a brief discussion of the frequency of magnetic shaping in real prePNe and PNe.

---

[1] Section 4 of that paper is a brief anecdotal exploration the influences of fields that motivated the present more extensive study.

## 2. Models and Results

*2.1 Methodology*

Aside from the field, the basic geometric, physical, and kinematic properties of the injected flow and the code used for the simulations are described in detail in previous papers in this series. All simulations are made using AstroBEAR (Cunningham et al., 2009ApJS..182..519C) in a $+x, +y$ quadrant on a 2-D grid of size $(\Delta x \times \Delta y) = (32 \times 64)$ kau using an adaptive-mesh-refinement ("AMR") method. The basic cells size is 500 AU (about the same as the grids used in earlier studies); however the cell width is dynamically decreased by $n$ powers of 2 when the state variables change rapidly. The toroidal field, $B_z$, is aligned to the $y$ axis (shown vertically). AstroBEAR uses a "2.5 dimensional" coordinate system, where the added dimension is azimuthal (the direction of the field). In the absence of photoionization radiative cooling is coronal (Dalgrano & McCray 1972ARA&A..10..375D). After running a variety of tests we find that, to all intents and purposes, the large-scale structure of the outflow after 600 y is realized with $n = 3$ (that is, cells of size $\delta r = 63$ au). However, some simulations that explore small-scale structures are run with $n$ up to 8 ($\delta r = 2$ au). We ran nearly 100 simulations in all, each with different combinations of the initial conditions of the injected flow.

A presumed steady, cold, diverging ("tapered"[2]), magnetized flow with an effective opening angle of 40° is injected into the grid normal to the surface of a spherical nozzle of radius $r_0$ of 1 kau. The injection speed of the magnetized wind is 400 km s$^{-1}$. This speed matches the axial expansion speeds of OH231.8+04.2 (Alcolea et al., 2001A&A...373..932A; Sánchez Contreras, 2018A&A...618A.164S) and CRL618 (Balick et al., 2013ApJ...772...20B; Riera et al., 2014A&A...561A.145R) and is similar to those in Hen2–104 (Corradi, et al., 2001ApJ...553..211C) and MyCn18 (Bryce at al., 1997ApJ...487L.161B, O'Connor et al., 2000ApJ...531..336O).

Other parameters are variations of tapered flow of the baseline model of Paper III. The tapered flow immediately encounters a stationary ambient medium (in practice, a much slower AGB wind (Ivezić & Elitzur, 2010MNRAS. 404.1415I) in which the form of the ambient density is $n_{amb}(r) = n_{amb}(r_0) (r_0/r)^2$, where $n_{amb}(r_0)$ is $2 \times 10^4$ cm$^{-3}$ unless noted otherwise. This value of $n_{amb}(r_o)$ was chosen to assure that the total mass of the AGB wind is about 0.7 M$_\odot$ (Höfner and Olofsson, 2018). We ignore photoionization and the inertial pressures of pre-existing equatorial disks. The latter, if present, have no major dynamical impact on the evolution of tapered flows.

Table 1. Model Parameters and Initial Values

| Parameters[a] | Initial Value |
|---|---|
| Attributes of the ambient gas at the nozzle surface | |
| density at nozzle surface $n_{amb,0}$ (cm$^{-3}$) | ($2\times10^4$, $2\times10^5$) |
| launch speed $v_{amb,0}$ (km s$^{-1}$) | static |
| temperature $T_{amb,0}$ (K) | 100 |
| 2-D grid quadrant boundary **($\Delta$x, $\Delta$y) (kau)** | 32×64 |
| Attributes of the magnetized flow at the nozzle surface on the $y$ axis | |
| nozzle radius $r_0$ (A.U.) | $10^3$ au |
| log$_{10}$ density $n_{flow,0}$ (cm$^{-3}$) | ($10^4$ to $10^7$) |
| injection speed $v_{flow,0}$ (km s$^{-1}$) | (200, 300, 400, 600) |
| log$_{10}$ toroidal field strength $B_{z,0}$ (Gauss) | ($10^{-2.5}$ to $10^{-0.5}$) |
| Gaussian 1/e taper angle $\phi_{flow}$ | 40° |
| temperature $T_{flow,0}$ (K) | 100 |
| injection duration $\Delta t_{flow}$ (y) | 600 |

*a* Bold: parameters that are specified in AstroBEAR simulations
*b* Values of $10^{37-39}$ g cm s$^{-1}$ are characteristic of fast winds (Bujarrabal et al., 2001)

The large-scale evolution of the nebula is completely dominated by ram pressure except in the immediate vicinity of shocks (e.g., Paper III, Fig. 3). Emission lines observed in prePNe arise solely in shocks (Raga et al. 2008A&A...489.1141R). Dust that is illuminated by and scatters starlight is a better optical tracer of the density distribution.

*2.2 Magnetized Flows*

*2.2.1. Summary of Non-Magnetic Flows.* Non-magnetic cases are standards of comparison for magnetic models. In this case the ram pressure of the field-free fast wind and the inertial resistance of the AGB wind shape the growing wind-blown lobe. Table 1 shows the adopted descriptors of the ambient environment and the injected flow used for the AstroBEAR simulations in this paper. Values in parentheses show the adopted ranges used in this study. The flows are "tapered" at the nozzle: i.e., values of the density, $n_{flow}(r_0)$ (abbreviated $n_{flow,0}$) and speed, $v_{flow,0}$ decline as Gaussians with polar distance from the y axis to their $1/e$ values at the "taper angle", $\phi_{flow}$. (The terms "opening angle" and "taper angle" are used interchangeably.) The subscript 0 refers to the values of these quantities at the intersection of the nozzle and the $y$ axis. For simplicity we have adopted $v_{flow,0} = 400$ km s$^{-1}$, $\phi_{flow} = 40°$, $T_{flow,0} = T_{amb,0} = 100$K, $n_{amb,0} = 2 \times 10^4$ cm$^{-3}$, and $T_{amb,0} = 100$K for nearly all of the simulations presented throughout this paper. "Lobes" are the bulbous regions of the outcomes. As shown in Paper III, $v_{flow,0}$ primarily determines the size of the lobe and $\phi_{flow}$ its aspect ratio.

---

[2] A "tapered" flow has a momentum distribution that declines with polar angle. We adopt the taper prescription of Lee & Sahai (2003ApJ...586..319L) in which the density and speed of the nuclear outflow are modulated from their axial values as a Gaussian with a "taper angle" denoting the angle at which the modulation is 1/*e*. See Ramos-Medina (2014apn6.confE..74R) for observations of the actual form of the taper in M1–92 (Fig. 1).

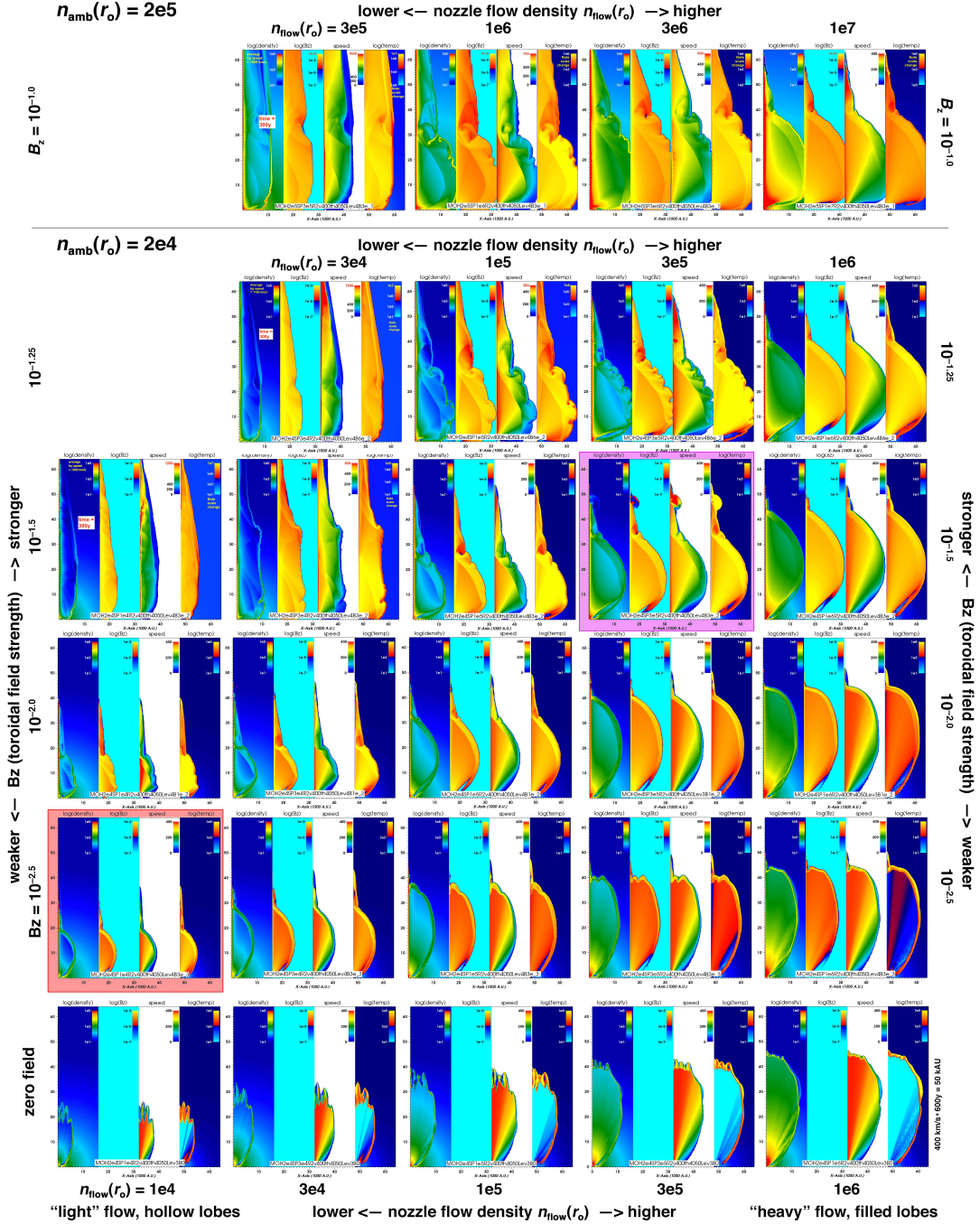

Fig. 2. An array of results from the simulations of steady stellar winds with toroidal magnetic fields at time $t = 600$y. Each panel contains plots of the logarithm of the density, the field strength, the flow speed (linear), and the local temperatures. The color scales are generally but not always exactly the same in each of the panels. The density contrast of the injected gas (see text) increases to the right and the strength of the toroidal field increases vertically. The bottom row shows examples of field-free models. The first of these is the baseline model of Paper III. The top row illustrates proportional changes in the ambient and flow densities and the magnetic field. Only the models in the left column have lobes that appear to be hollow. The models shown in red and magenta boxes are the light-flow "baseline model" and the heavy-flow "reference model", respectively (see the text).

We show the outcomes of five field-free simulations spanning a range of "density contrasts", $0.5 \leq n_{flow,0} / n_{amb,0} \leq 50$, in the bottom row of Fig. 2. As discussed in Paper III, a density contrast < 1 produces a "light" flow in which the size of the lobe is sensitive to the inertial pressure of the ambient medium; i.e., the mass of displaced gas (left panels). The resulting lobe is hollow and its expansion speed is a fraction of the injection speed of the wind. "Heavy" flows (right panels) are less affected by the ambient gas; they penetrate faster and further into the ambient gas than light flows. The lobes formed by very heavy winds are likely to appear filled.

In nearly all of the figures the inner edges of lobe walls are formed by a fast reverse shock where the flow streamlines intersect the walls of slower displaced gas obliquely. The outer perimeters of the walls are bounded by a slower forward shock where thermal pressure near the wall periphery pushes laterally into the ambient gas. As discussed in Paper III, thin-shell instabilities can cause the leading edges of the lobe to corrugate. The speeds of both shocks usually reach their maxima where they cross the y-axis.

*2.2.2. Models of Fast Magnetic Flows.* The upper rows of Fig. 2 show the results of the models with toroidal fields of various strengths $B_z$ injected into the flow at the nozzle. The density contrasts are the same within each column. The elapsed time is $t = 600y$ with a few exceptions for very rapidly growing lobes in the upper left panels. We will refer to two of the models, the "light-flow baseline model" and the "heavy-flow reference model" throughout the paper. These are portrayed with red and magenta backgrounds, respectively, in Fig 2. The baseline model is a typical light flow model (density contrast $n_{flow,0} / n_{amb,0} = 0.5$) and a modest magnetic field ($B_{z,0} = 10^{-2.5}$G). The reference model (magenta box) illustrates a heavy flow, $n_{flow,0} / n_{amb,0} = 15$, with a stronger field, $B_{z,0} = 10^{-1.5}$G. Both models differ mostly in scale.

The outcomes illustrate how the sizes, shapes, and wall thicknesses of the lobes vary with $n_{flow,0} / n_{amb,0}$ and $B_z$. Except in extreme cases (light, highly magnetized flows), the lobes remain easily recognized by their walls, bounded as usual by the shocks formed at the terminus of the free-flowing nuclear winds and at lobe interface with the slower ambient gas. Note also that the walls of lobes are thickened by the internal pressure of compressed magnetic fields, especially for light flows. Obviously most of the thin-shell instabilities are suppressed and the lobes become increasingly closed as the walls thicken.

The most obvious feature produced by the toroidal field is an axial protrusion beyond the lobe's leading edge. The protrusions and the bow shocks that trail its leading tip are distinctive signatures of all models of magnetized flows. Such protrusions—discussed in more detail later—are present in some fashion in all magnetized simulations in Fig. 2, including the extreme cases in the upper left portions of the figure. These are created by a very thin and highly magnetized jet that forms on the symmetry axis starting immediately above the nozzle. In most of the models of Fig. 2 jets have a higher speed of growth than the lobes so they steadily pull away from the lobe.

The trends with increasing $B_z$ are clear in in Fig. 2. The length of the protrusion grows in proportion to the lobes. The thicknesses of the walls of the lobes increase as the fields compressed in their interiors rise. This is most obvious in fourth column in Fig. 2 in which $n_{flow,0} = 10^{5.5}$ cm$^{-3}$. (The lobe walls become crenellated at large $B_z$; these ripples are unrelated to shear flows).

Extreme cases are found in the upper left of Fig. 2. The panels with $(B_z, n_{flow,0}) = (10^{-1.5}, 10^{4.0})$, and $(10^{-1.25}, 10^{4.5})$ illustrate how the speed of the jet is greatly enhanced as the walls of the flow thicken and begin to impinge on the symmetry axis. (These panels are shown after only 300 yr when the tip of the jet leaves the top of the grid.) In both examples the inertial pressure of the ambient density cannot contain the combined ram and magnetic pressures. The internal temperature of the lobe and protrusion exceed $10^7$K and the growth speed of the lobe tip (seen in the legends of the speed panels) exceeds 1000 km s$^{-1}$. We know of no obvious examples of such "flame throwers" among prePNe, though Hen2-437 (Fig. 1), Hen3–401, and IRAS 13208-6020 bear some resemblances. (These prePNe should be observed more thoroughly to test this prediction.)

Furthermore, note that the gas temperatures inside the lobe walls are often in excess of $10^{4.5}$ K in the upper rows of Fig. 2. At this temperature electron collisions can excite line emission and partially ionize the elements such as N, O, and S that emit visible forbidden lines. This provides an alternate explanation to wind shocks for the line emission seen in lobe walls in the lower row of Fig. 1.

Understanding the growth of prePNe is a matter of following the pressures that govern their evolution. In Fig. 3 we see that ram pressure generally dominates the structure of the growing features, even in the lobe walls and the jets where the magnetic field is most compressed. In addition, the ratio of magnetic to thermal pressure is ≈1.5 except very locally in supersonic shocks. Thus the pattern of local temperature closely follows that of the field strength throughout the lobe.

It is interesting to note that the shapes of the models (and the ratios of the lengths of the protrusions and lobes) are nearly the same along +45° diagonals in Fig. 2. Models with the same ratio of magnetic and ram pressures, $[B_z^2/8\pi] / [½ \rho v^2]$, lie close to these diagonals. Thus, for a specific measured expansion speed, ratios of the protrusion and lobe lengths identify families of outcomes with similar densities and fields.

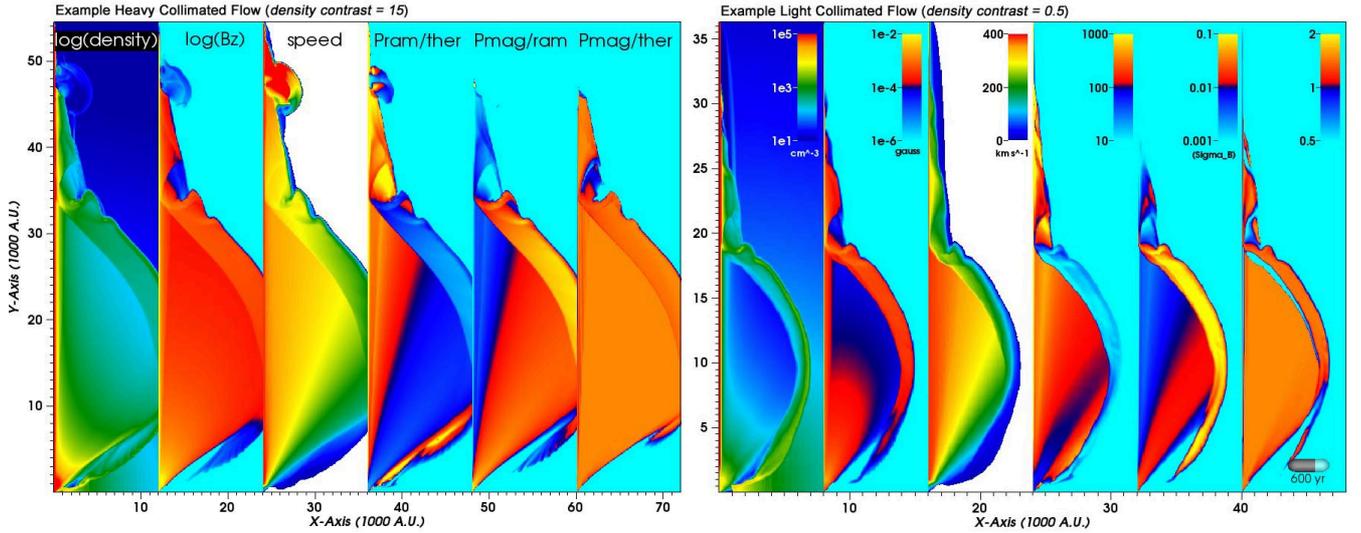

Fig. 3. Two sets of panels with outcomes for the reference (heavy flow; left) and baseline (light flow; right) models at $t$ = 600y. The first three panels show the flow density, field strength, and speed distributions. The last three panels show the ratios of ram, thermal, and magnetic pressures that influence the evolving structure of the lobes. The ambient density, flow speed, and opening angle are fixed at $2 \times 10^4$ cm$^{-3}$, 400 km s$^{-1}$, and 40°, respectively, at the nozzle. Note the differences in scale sizes between the frames.

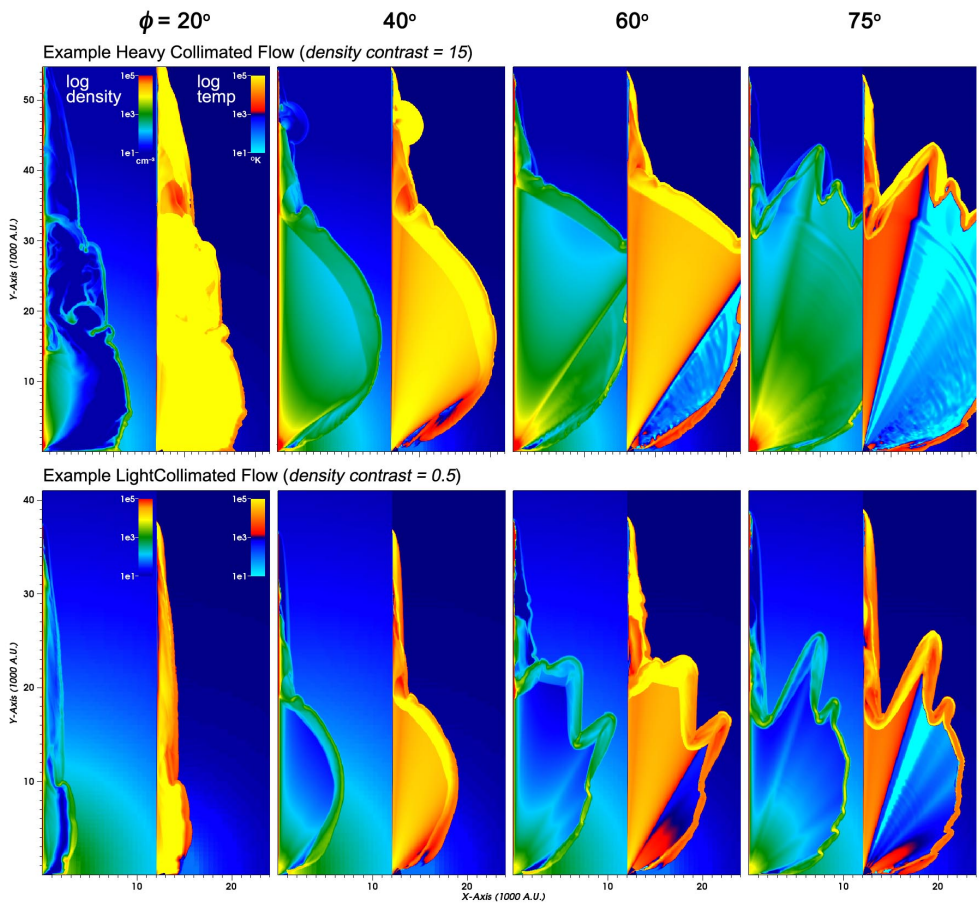

Fig. 4. The impacts of changes in the opening angle of the injected flow on the 600-y outcomes of the simulations for the heavy reference model (top row) and the light baseline model (bottom row). The ambient density, $2 \times 10^4$ cm$^{-3}$, flow speed, 400 km s$^{-1}$, are fixed. The opening angles of the flow at the nozzle in the four panels, $\phi_{\text{flow}}$, are 20°, 40°, 60°, and 75° from left to right. Density and temperature distributions are shown side by side. The scale size of the bottom row is 150% that of the top row. The color legends are the same throughout. Note the consistency of the lengths of the jet and the lobe in each row of the figure.

*2.2.3. Impacts of Parameter Changes.* We now explore how changes in flow parameters affect large-scale model outcomes.

1. *Density contrast.* Density contrast is a vital model parameter that was discussed when Fig. 2 was presented. In that discussion the density of the ambient gas at the nozzle was fixed at $n_{amb,0} = 2 \times 10^4$ cm$^{-3}$. The top row of Fig. 2 shows the effects of increasing $n_{amb,0}$ by a factor of ten. The outcomes are very similar to those shown immediately below them (where the density contrast, $n_{flow,0} / n_{amb,0}$ is the same), just as in the non-magnetized case (papers II and III).

2. *Flow Speed.* Changes in flow speed effectively set an overall scale factor for the both the jet and the lobe for $200 \leq v_{flow,0} \leq 600$ km s$^{-1}$. Shapes are only slightly affected for the light "baseline" and heavy "reference" flows of Fig. 2. This is a direct result of the overwhelming large-scale dominance of ram pressure (Fig. 3).

3. *Opening angle.* Different opening angles produce a range of shapes, as illustrated by two specific cases in Fig. 4. In all cases the density of ambient gas at the nozzle and the injection speed are both fixed at $n_{amb,0} = 2 \times 10^4$ cm$^{-3}$ and $n_{flow,0} = 400$ km s$^{-1}$, respectively. The opening angle increases from left to right. The results for a heavy flow case with a moderate field strength, $(B_z, n_{flow,0}) = (10^{-1.5}, 10^{5.5})$, are shown in the top row, and for a light flow and low field strength, $(B_z, n_{flow,0}) = (10^{-2.5}, 10^{4.0})$, appear in the bottom row. Fig. 4 illustrates that the lengths of the jets change only marginally from left to right; however, the dense lobe walls vary strongly in shape, size, and temperature.

## 2.3 Jets

Jets arise in every model in figs. 2, 3, and 4. They are the most conspicuous signature of all magnetized models. Hoop stresses in the flow quickly deflect the injected gas towards the symmetry axis forming a very thin and dense column, or "jet", adjacent to the *y* axis. The internal speed of the jet is close to that of the flow at the nozzle, ≈400 km s$^{-1}$ (with the exception of the flame-thrower examples mentioned earlier).

For steady-flow injection at the nozzle one might expect the column to also be of uniform density, like a steady jet. However, the density profile along the jet in the heavy reference model (and most of the other simulations of heavy flows) increases parabolically as $A(r/10^3 au)^2$ along its length, where $A = 2.0 \times 10^7$ cm$^{-3}$, at all times starting within 100 years of launch. After $t = 600$y the peak density of the jet is $4.0 \times 10^{10}$ cm$^{-3}$ located at $4.3 \times 10^4$ au from the star. The peak increases in density during its propagation. (N.B.: if the field is ten times weaker, $B_{flow,0} = 10^{-2.5}$ G, then A is ten times smaller).

The dense peak of the jet moves (and will continue to move) ballistically through its surroundings where the ambient density is only ≈10 cm$^{-3}$ and dropping rapidly. Were it fully visible, the jet would resemble a dense knot, or "bullet", at the jet tip with a short tail and accompanied by a leading bow shock of displaced ambient medium.

Of course, the bullet-dominated density profile of the jet will be affected if the mass injected by the stellar flow at the nozzle isn't steady. That is, a brief injection will form a dense, neutral magnetized, magnetically self-contained bullet. Intermittent or periodic accretion rates on time scales as a stellar merger or periastron passage of a companion mass donor may lead modulated density profiles with much the spatial cadence. Similarly, changes in the orientation of an accretion disk will produce jets that wobble. (See models of collimated mass flows from close binary stars in eccentric orbits by Velázquez et al., 2014ApJ...794..128V and Riera et al., 2014A&A...561A.145R.)

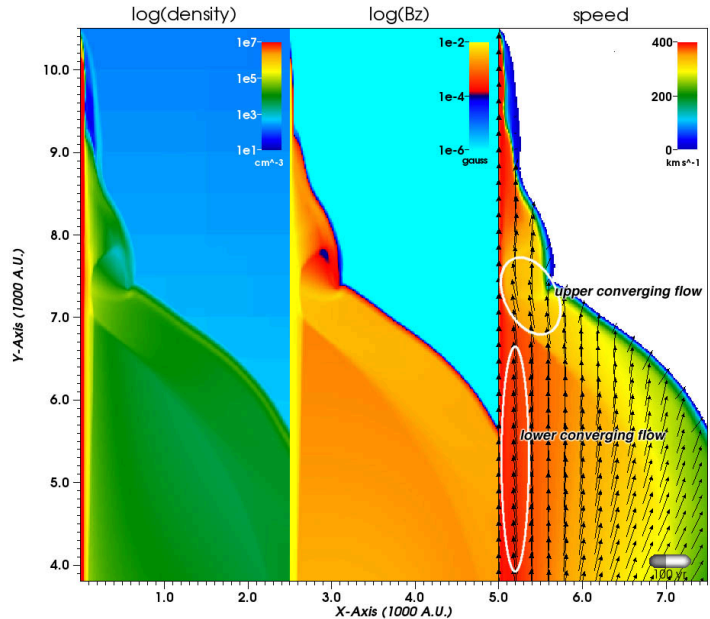

Fig.5. Left: moderate-resolution ($\delta r = 16$ au) simulations of the density distribution, (left panel), toroidal magnetic fields ($B_z$, left), and flow patterns (right) for the reference model at $t = 100$y. Only the upper portions of the simulations are shown for clarity. The color legends are the same as those in Fig. 3 in order to facilitate comparisons. The flow vectors are all the same length. The jet is easily recognizable along the *y*-axis of the first two panels. Streamlines converge towards the jet in two zones outlined in white.

*2.4 The Formative Years*

Jets form nearly immediately in our simulations. A very high-spatial resolution model (not shown) was run for the first 20 y ($\delta r$ = 2 au; 0.2% of the launch radius). The jet is already obvious at $t$ = 10y with an initial radial speed 390 km s$^{-1}$, essentially that of the injected tapered flow along the *y*-axis. After 20 y we began to decrease the spatial resolution of the grid to $\delta r$ = 4 au in order to follow the growth of the structure. Fig. 5 ($\delta r$ = 16 au) shows the result at $t$ = 100 y. In this case the speed of the dense leading knot at the tip the jet increased to 430 km s$^{-1}$ as the toroidal field diverted more of the streamlines towards the axis in two locations, as seen in the circled zones in the right panel of Fig. 5. The upper convergence zone—where the local field is strongest—is the most influential one. Note that at $t$ = 600 y the tip speed remains steady whereas speed of the upper edge of the lobe behind it has slowed to 280 km s$^{-1}$. This sort of behavior is characteristic of most of magnetic models shown in Fig. 2.

If accurate, the early results constrain detailed but usually short-duration models of outflows from of common envelopes and accretions disk (see the review by Frank et al. 2018Galax...6..113F).

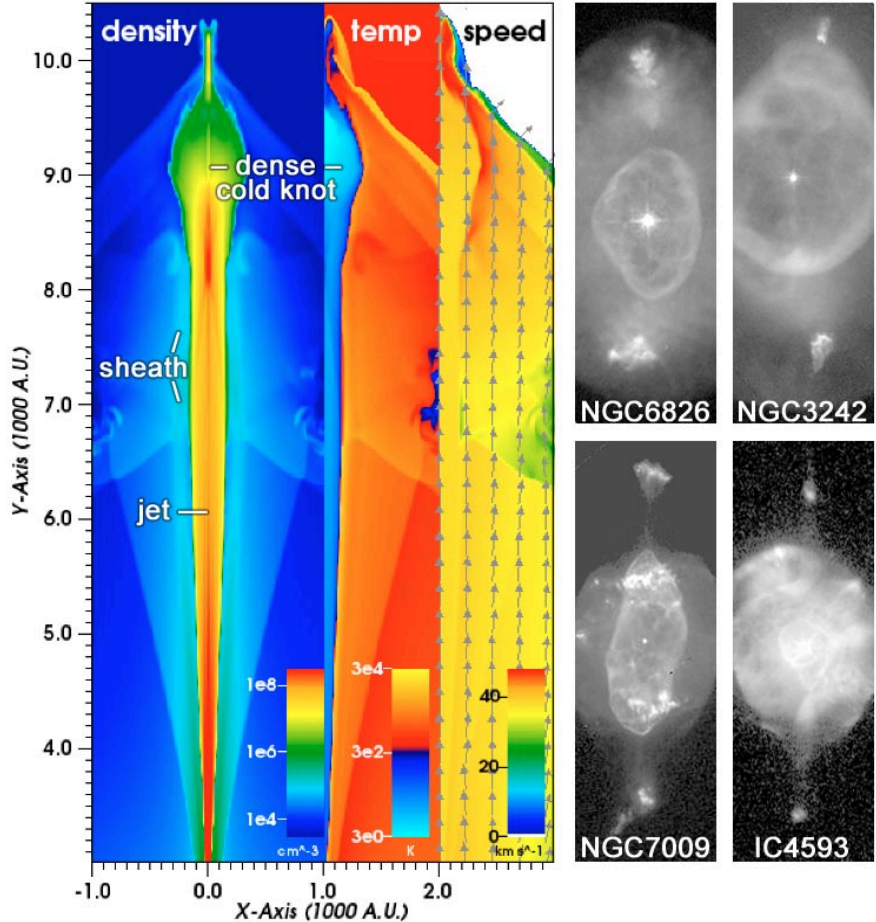

Fig. 6. Left: portions of the outcome of a simulation at $t$ = 1000 y in which the density contrast at the top of the spherical nozzle is 15, $B_z$ = $10^{-2.5}$ G, the injected wind speed is 40 km s$^{-1}$, and other flow parameters are those in Table 1. Right: high-quality [NII] images of prominent FLIERs in four fully ionized PNe. See section 3 for a comparison of the model and the images.

*2.5 Slow Injections*

Guerrero et al., 2019 (preprint) have compiled statistics on the sizes, (Doppler) expansion speeds, and the kinematic ages for 57 ionized PNe that show signs of collimated outflows. Some 15% of the nebulae fall into the "high-speed group", including all of those identified in section 2.1 that motivated this paper. The remaining 85% form a "low-speed" group characterized by a mean expansion velocity of 64 km s$^{-1}$. The mean nebular age of the latter group is 2000 yr.

Accordingly, we have run a limited set of simulations of models with injection speeds $v_{\text{flow},0}$ reduced by a factor of ten in order to explore whether the outcomes of the slow injected winds show substantial differences from the fast injection models above. We chose sets of density contrasts and the ratio of magnetic to thermal pressures at the nozzle that are the same as in the fast models presented earlier. That is to say that we decreased both $v_{\text{flow},0}$ and $B_z$ by a factors of ten. We also retained the mass of the surrounding ambient gas from the AGB wind; that is, $n_{flow,0}$ remained $2 \times 10^4$ cm$^{-3}$. The results of the "slow-injection" simulations were computed for a duration of 1000 y; however, after about 500 y the growth pattern is ballistic.

Slow-injection models with $v_{flow,0}$ = 40 km s$^{-1}$ were run for all combinations of two density contrasts, 15 and 0.5, and two values of $B_z$, $10^{-3.5}$ and $10^{-2.5}$ G (these results are not shown). Aside from a spatial scale factor of ten the outcomes largely resemble the corresponding fast-injection cases. Specifically, the lobes retain their previous shapes. However, Fig. 6 demonstrates that there are some noticeable changes in the structures of the jets and their bow shocks. The jets are stubbier (though they still protrude beyond the front of the lobe), their tips can be more bulbous (larger cross sections), considerably colder, and, by $t$ ~1000 y, the proper motions of the knots at their tips are 10–20% slower than the gas in the magnetized sheath that surrounds them. The bow shocks are more prominent than earlier since the local ambient density is ~$10^2$ times larger than before.

In section 3.1.2 we will compare the model of Fig. 6 to some well-observed PNe that contain pairs of axial knots.

*2.6 Limitations and Reliability*

One significant limitation of the present simulations is the 2.5-D grid that we used in order to be able to expediently run over 80 different models. By confining ourselves to 2.5-D models we are only able to focus on toroidal fields, $B_z$, in which the radial and azimuthal components of the field, $B_r$ and $B_\phi$ respectively, are unimportant for global shaping[3]. This is not a serious limitation since $B_r$ falls off much more rapidly with distance than $B_z$.

We emphasize that the adjacent dense sheath threaded by toroidal fields surrounds the jet and inhibits and confines any instabilities in the present 2.5-D models. However, we cannot fully realize the kink instabilities of various types that normally arise in a full 3-D treatment. According to the standard Kruskal–Shafranov criterion for a magnetized jet of radius $r_j$ the kink instability will grow if

$$\left|\frac{B_\phi}{B_z}\right| > |(\beta_z - 1)kr_j|$$

where $\beta_z \propto P/B_z^2$, P is the local pressure, and $k$ is the characteristic wavelength of the current-driven perturbations (Boyd & Sanderson 2003). The above equation demonstrates the well-known capacity of a toroidal field to behave as a "spine" that stabilizes the flow. As we will discuss in section 3.2, kink modes in published 3-D jet models associated with bipolar outflows need not be disruptive to global flows. Even so, it is not clear if such conditions will obtain in the general case (i.e., in real prePNe or PNe) since the stability of the jets will depend critically on the details of the nature of the jet launching.

Finally, we are confident of the fidelity of the large-scale lobe dimensions, wall shapes and thicknesses, and expansion speeds by comparing models with identical inputs but very different spatial resolutions. We found that δ$r$ can be doubled every ≈50 y without significant impact on the large-scale shapes of the lobes. Monitoring the jet in detail is not necessary for "large-picture" simulations that can be compared to observations.

**3. Discussion**

*3.1 Comparison of Outcomes with Images*

*3.1.1. PrePNe.* We have carefully compared our various simulations to HST images of prePNe, including those in Fig. 1. We exploited various catalogues of prePNe or very compact PNe (Sahai & Trauger 1998AJ....116.1357S, Sahai et al., 2011AJ....141..134S and Stanghellini et al., 2016ApJ...830...33S) a well as sundry publications on individual prePNe for clear signs of protrusions and fast bullets near the leading edges of the flow. The objects in Fig. 1 represent a majority of all of the candidates that we found.

None of these knots and bow features is explained by other formation paradigms. The first four images in both rows of Fig. 1 best the show axial protrusions that we have identified with invisible axial jets. The shapes of K3–35 and M1–92 are comparable to the reference model mentioned earlier and its shape family with similar values of $\sigma_B$. M1–339 may be a case of a wide opening angle flow where $\phi \gtrsim 60°$. M1–31, Hen2–115, M1-12, He2-320, and IRAS19475+3119 (not shown) all have conical protrusions with sharp tips emanating from complex lobes with thin walls, reminiscent of bow shocks from the tips of invisible axial jets. Light-flow simulations mimic their structures especially well. He2-430 has long, thin, and hollow protrusions and no apparent lobes. Light-flow "flame thrower" simulations seen in the models of Fig. 2 may be counterparts. He3–1475 and CRL618 contain visibly shock-excited knots with trailing bow shocks (Balick et al., 2014ApJ...795...83B; Fang et al., 2018ApJ...865L..23F) that are good candidates for fast bullet-tipped outflows (Riera et al., Velázquez et al., 2014ApJ...794..128V) that we discussed in section 2.3. The high proper motion knots of both CRL618 and Hen3-1475 are bright in the line of [Fe II]λ1.64μm (Balick et al., 2013ApJ...772...20B; Balick, unpublished)—a certain sign of strong and fast shocks (Paper II). IRAS16594-4656 and IRAS19475+3119 have similar knots with bow waves that are seen in images of reflected starlight.

---

[3] One significant limitation of the present simulations is the quasi-planar 2.5-D grid that we used to run over 80 different 2.5-D models. The utility of AMR decreases as the grid tends to fill up with finest level grids as the nebula fills larger volumes of the computational space. This renders 3-D models impractical for multi-parameter variation studies such as those presented here.

*3.1.2. Mature (Ionized) PNe.* Symmetric pairs of relatively dense low-ionization knots (sometimes stubby jets) lie along the major axes of many mature PNe (Gonçalves et al., 2001ApJ...547..302G). Among these (in random order) are Hen2–104 (where the axial knots are shock excited), 141, and 186; Hen3–1312; Fleming 1; Hubble 4; IC 2149, 4593, 4634, and 4846; Jonkere 320; K3–36 and 4–47; M1-16, 66 and 73, M2–46; M3–29; NGC 3242, 3918, 6210, 6337, 6543, 6751, 6826, 6881, 6905, 7009[4], 7354, and 7662. It is quite clear that the small axial knots and stubby jets are dense and centrally launched. We investigate whether they are the persistent remnants of neutral bullets formed by toroidal fields during their prePN phases.

In Fig. 6 we showed that toroidal fields are capable of producing dense neutral knots whose speeds are about the same as the speed of the gas injected at the nozzle. The very high density of the knots assures that their interiors remain cold and neutral for at least 1000 y as the knots move through the surrounding ionized gas near their path. Thus many or most of the knots and jets may survive as distinct low-ionization structures as the the central star ionizes the rest of the nebula.

Four of the best examples of pairs of dense axial knots are shown on the right side of Fig. 6. The toroidal field model to their left demonstrates the success of the toroidal field simulations to account for their outstanding features. (No other mechanisms have been proposed.) The are probably formed *in transit* since detailed hydrodynamic models of accretion disks and common envelopes in the core (section 1) do not predict their formation.

The knots in Fig. 6 are seen in [NII] emission. Spectroscopic observations (Balick et al. 1993ApJ...411..778B and 1994ApJ...424..800B) show that they are bright in [NII] and collisionally excited lines of lower ionization. However, there is no corresponding [OIII] emission, so the hot central star that photoionzes $O^+$ elsewhere in the nebula is not affecting the ionization of the gas in these knots. Instead, the emission is either the result of shocks or the result of local magnetic turbulence and coronal (thermal) excitation. Many spectroscopic studies of other specific low-ionization knots (Gonçalves et al. 2009MNRAS.398.2166G; Corradi et al., 1999ApJ...523..721C; Corradi et al., 2000ApJ...535..823C) have concluded that the knots are shocked; however, shock models by Raga et al. (2008) require a very narrow range of shock speeds, 80–100 km s$^{-1}$, or else the [OIII] line will also be readily observable. This seems highly unlikely in every case. Moreover, Guerrero et al., 2019 find that shock speeds of 80–100 km s$^{-1}$ are factors of two to four larger than the measured Doppler speeds observed for the low-ionization knots in all four of our examples. As a result, the evidence argues that the knots are not ionized and excited by shocks. Instead, the knots may be coronally excited, as predicted by the temperatures in most of our models.

*3.2 Comparison with Previous Models*

Dennis et al. (2009ApJ...707.1485D) have succinctly reviewed the history of MHD simulations of stellar outflows. The concept of toroidal fields in stellar winds (in their case fast isotropic winds) shaping lobes as they encounter a spherically symmetric external density with axial protrusions dates back to Różyczka & Franco (1996ApJ...469L.127R). Like others to follow (e.g. Frank et al., 1998ApJ...494L..79F), they find that the growing wind-blown bubble (which appears as a cavity in optical images) develops axial protrusions as streamlines are magnetically redirected to the symmetry axis by hoop stresses.

The presence of hoop stresses around the symmetry axis of the magnetized flow is a common ingredient in all magnetized outflow models. So all models predict a jet-like axial flow with a possible bow shock of displaced ambient material. Thus the applicability of specific models to actual images must be decided on other more subtle indicators. This is a difficult task without exquisite observational data.

The most analogous simulations to those presented here were published by García-Segura, et al. (1999ApJ...517..767G; hereafter "G-S+99") and Steffen et al. (2009ApJ...691..696S; hereafter "S+09"). As we noted in section 1, all of these share the same and highly specific flow source function: magnetically infused isotropic winds that are centrifugally launched from a rotating AGB star or white dwarf with embedded toroidal fields. The authors use this paradigm to fix the speeds, geometries, and mass influx rates of the flows that enter the grid. (We make no such assumptions). The nozzle and their grid resolution are ≈500 au (our largest grid) for their fully three-dimensional simulations using the MHD code ZEUS-3D.

The "magnetic tower" simulations of Huarte et al. (2012ApJ...757...66H) are of particular interest for this study because of the fate of the toroidally dominated jet which formed in their models. A key difference between our models and those of Huarte et al. were that their bipolar outflows were Poynting flux dominated (meaning they were driven entirely by magnetic pressure rather than ram pressure) from an injected flow. In spite of this difference, the toroidal fields in their simulations still squeezed plasma towards the axis forming ram pressure dominated collimated jets. Eventually these jets experienced strong kink instabilities that led to the jet beam breaking up into a series of knots with complex field topologies. What is noteworthy is that although the jets were fragmented, the overall collimation of the jet flow was not changed. The individual knots continued to propagate in the polar direction. Most importantly, the shape of the bipolar lobes was not strongly affected by the changes in the jet flow within them. The configuration explored in Huarte et al. was also reproduced in High Energy Density Laboratory Astrophysics (HEDLA) plasma experiments (Ciardi et al 2007PhPl...14e6501C, 2009ApJ...691L.147C) in which the kink-mode fragmentation of the internal jet was observed as well as the persistence of the flow collimation.

---

[4] NGC 7009 is a notable but rare example of an ionized PN that may have been shaped by a large-scale toroidal field (Steffen et al., 2009RMxAA..45..143S) or by a surrounding poloidal field (Albertazzi et al., 2014Sci...346..325A).

The present outcomes share many similarities with those published by G-S+99 despite our disparate methodologies, flow geometries and inflow speeds. The mass injection parameters that they adopted for their magnetic simulations for AGB winds (labelled M through U) are found in their Tables 1 and 2. These values correspond roughly to our parameters $n_{flow,0} \approx 2 \times 10^6$ cm$^{-3}$ and $v_{flow,0} = 10$ km s$^{-1}$ at the nozzle surface at $r_0 = 500$ au. That is, their value of $n_{flow,0}$ is about 100 times larger and their and $v_{flow,0}$ is 40 times smaller than ours (depending on the value of $\Omega_{AGB}$ that they adopted). Our momentum injection rates, $\rho v$, are comparable, so we agree on the growth speeds of the lobes (though not the jets). G-S+99 adopted $\sigma_B = 0.01$, slightly smaller the values that we used at the base of the wind. They also added a source of UV photons that both ionize and heat the gas.

Similarly, the outcomes found by S+09 show many resemblances to ours. S+09 specifically study the flow from a hot white dwarf that ionizes its surroundings into an ancestral AGB wind. They used much the same methodology as G-S+99 used, though their mass injection rates, and speeds, and momentum fluxes are somewhat different. One of the principal differences of their models with the others is the growth rate of the lobes and jets, $\approx 50 - 70$ km s$^{-1}$ after $\approx 2000$ y. This is much smaller than observations show. This probably stems from the small mass injection rate that they assumed, $10^{-8}$ M$_\odot$ y$^{-1}$. That is, the injected winds are so "light" that their growth is highly retarded by the ambient gas that they encounter and displace.

The simulations by Frank et al. (2004ApJ...614..737F) and Dennis et al. (2009) considers an external but local poloidal field in the winds of an external, extended, rotating accretion disk, not surface of the star itself. This paradigm is widely applicable to young stellar objects. The field emanates from a rotating disk through which purely radial (untapered) streamlines from the nucleus flow at an angle to radial, non-magnetized winds from the star. Both winds have differing mass loss rates, speeds, and geometries. Magnetic tension in the rotating field source is transmitted into the combined flow whose geometry is modified out to a distance of the Alfvèn radius, at which point the magnetic tension of the disk wind creates local turbulence. Further downstream the composite field becomes largely toroidal, so the subsequent flow has most of the same attributes as those in most studies: an axial jet with a narrow conical bow shock.

Ciardi, et al. (2013PhRvL.110b5002C), Albertazzi et al. (2014Sci...346..325A), and van Marle et al., (2014A&A...570A.131V) examined how external poloidal fields (parallel axial field lines) affect the shapes of PNe formed by steady isotropic fast winds from a central star. This paradigm is not the one adopted here, so the similarities in the outcome shapes seems to be largely coincidental.

## 4. Summary and Conclusions

The primary goal of this paper is to determine whether a significant subclass of prePNe with leading axial protrusions (e.g., M1–92) and high-speed knots (e.g., CRL618) can be successfully simulated so that their evolution can be traced back in time. We have calculated an array of 2.5-D models of flows with embedded toroidal magnetic fields until the resulting lobes and jets grow homologously (elapsed times > 500 y). All of the simulations assume a tapered flow from the nozzle that is launched into the (presumed static) remnant of a previously steady and isotropic AGB wind. We varied the density, speed, and opening angle of the tapered flow to generate a series of outcomes shown in figs. 2 through 6. The models span a range of field strengths and flow densities (more precisely, ranges of magnetic-to-ram pressure ratios at the nozzle), fast and slow injected winds, and flows with various opening angles. In every case—and as expected—magnetic hoop stresses deflect the wind streamlines towards the field symmetry axis forming dense axial jets and knots that overtake the leading edges of the lobes behind them. These magnetically induced flows invariably produce the protrusions and associated bow shocks much like those observed leading edges of wind-blown lobes. Indeed, these protrusions are unique to magnetic models. They render the successful identification of magnetically influenced objects straightforward (Fig. 1).

The shapes of features formed by collimated stellar outflows result from an evolving combination of ram, magnetic, and thermal pressures. Ram pressures dominate over the magnetic and thermal pressures everywhere for outcomes that best match the images in Fig. 1. Nonetheless, magnetic pressures create local hoop stresses near the symmetry axis that deflect nearby flow streamlines onto the axis of the toroidal field to create a thin, cold, and invisible jet, often with knots of sharply increased density at their leading edge of the jet. In other flow directions the field is compressed into the walls of the lobe. The magnetic pressures within the walls generally thicken and heat them to temperatures ≥30,000K where emission lines of low-ionization species are thermally excited (rather than shock excited).

Upon launch the (presumed steady) flows encounter a much slower AGB wind downstream whose total mass is about 0.7 M$_\odot$ (i.e., whose density is $2 \times 10^4$ cm$^{-3}$ at the nozzle surface of radius 1000 AU.) The panels of Fig. 2 are representative of the range of shapes and sizes of the structures formed by toroidally magnetized winds injected from the nozzle at 400 km s$^{-1}$ after 600 y (after which they grow homolgously). In this figure the field strength and density contrast are varied to show the full range of lobe and protrusion shapes that fields are likely to produce. The shapes fall into families of very similar characteristics defined approximately by $\sigma_B$, the ratio of the magnetic energy density to the ram pressure at the nozzle surface. Thus this ratio can be measured from images. We repeated some of these simulations with a much slower injection speed of 40 km s$^{-1}$. The size scales change but the shapes remain basically the same.

After 500 y the tip of the protrusion moves ballistically into the sparse slow AGB winds downstream. This means that speeds and the kinematic ages of the protrusions are reliable estimates of the true injection speed, $v_{\text{flow},0}$, and kinematic ages, $\theta/\dot{\theta}$, where $\theta$ is the full angular length of the tip of the protrusion. (This is a result that does not generally apply to light flows without magnetic fields, cf. Paper III.) Finally, for fixed $\sigma_B$, the opening angle of fast winds determines the shapes of the lobes but has very little effect on the length or speed of magnetically formed axial structures.

We showed in Fig. 6 that very dense axial knots formed in slow, heavy flows will survive and can eventually become FLIERs or low-ionization knots once the their surfaces start to become ionized by stellar UV, shocks, or coronal heating. Their high densities and related UV opacities assure that their interiors remain neutral and cold. However, observations show that the speeds of the FLIERs and knot are nearly always less than ~40 km s$^{-1}$ (Guerrero et al., 2019). Therefore it seems improbable that shocks can excite the optical forbidden lines such as [NII] that are observed in most all cases.

We conclude with an estimate of the importance of nebular shaping by toroidal fields. Morphological evidence of magnetic shaping—that is, leading protrusions—occurs in about 20% of the prePNe for which good HST images are available. Therefore toroidal magnetic fields, even if they are transitory—probably influence the evolution of a significant fraction of prePNe and PNe early in their evolution.

**Acknowledgements**

It is a pleasure to thank Martin Guerrero for a critical and very helpful reading of an early version of this paper. His comments vastly enriched parts of sections 3 and 4. We are grateful to Jonathan Carroll-Nellenback and the many people in the Physics-Astronomy Computational Group at the University of Rochester who developed and maintain the Astrobear code. We gratefully acknowledge a decade of valuable discussions with Noam Soker.

We acknowledge support from program AR-14563 that was provided by NASA through a grant from the Space Telescope Science Institute, which is operated by the Association of Universities for Research in Astronomy, Inc., under NASA contract NAS 5-26555. This paper is partially based on observations made with the NASA/ESA Hubble Space Telescope, obtained from the MAST Archive at the Space Telescope Science Institute, which is operated by the Association of Universities for Research in Astronomy, Inc., under NASA contract NAS 5-26555. Support for MAST for non-HST data is provided by the NASA Office of Space Science via grant NAG5-7584 and by other grants and contracts.